\documentclass[manuscript]{aastex}

\usepackage{epstopdf}

\tighten
\newcommand\ph{\phantom{0}}
\newcommand\be{\begin{equation}}
\newcommand\ee{\end{equation}}

\newcommand\febyh{\ensuremath{[\mathrm{Fe}/\mathrm{H}]}}
\newcommand\msun{\ensuremath{M_\odot}}

\newcommand\SC{\ensuremath{\mathrm{16\,Cyg}}}
\newcommand\SCA{\ensuremath{\mathrm{16\,Cyg\,A}}}
\newcommand\SCB{\ensuremath{\mathrm{16\,Cyg\,B}}}
\newcommand\SCAB{\ensuremath{\mathrm{16\,Cyg\,A\ and\ B}}}

\begin{document}
\shortauthors{Verma et al.}
\shorttitle{Helium abundance in \SC}

\title{Asteroseismic estimate of helium abundance of a solar analog binary system}
\author{Kuldeep~Verma\altaffilmark{1}, 
Jo\~ao~P.~Faria\altaffilmark{2,3},
H.~M.~Antia\altaffilmark{1},
Sarbani~Basu\altaffilmark{4}, 
Anwesh~Mazumdar\altaffilmark{5},
M\'ario~J.~P.~F.~G.~Monteiro\altaffilmark{2,3},
Thierry~Appourchaux\altaffilmark{6},
William~J.~Chaplin\altaffilmark{7,8}, 
Rafael~A.~Garc\'{i}a\altaffilmark{9},
Travis~S.~Metcalfe\altaffilmark{10,8}
}
\altaffiltext{1}{Tata Institute of Fundamental Research,
Homi Bhabha Road, Mumbai 400005, India}
\altaffiltext{2}{Centro de Astrof\'{i}sica da Universidade do
Porto, Rua das Estrelas, 4150-762 Porto, Portugal}
\altaffiltext{3}{Departamento de F\'{i}sica e Astronomia, 
Faculdade de Ci\^encias da Universidade do Porto, Rua do 
Campo Alegre, 4169-007 Porto, Portugal}
\altaffiltext{4}{Astronomy Department, Yale University, P. O. 
Box 208101, New Haven, CT 065208101, USA}
\altaffiltext{5}{Homi Bhabha Centre for Science Education,
TIFR, V. N. Purav Marg, Mankhurd, Mumbai 400088, India}
\altaffiltext{6}{Institut d'Astrophysique Spatiale, Universit\'{e}
Paris XI-CNRS (UMR8617), Batiment 121, F-91405 Orsay Cedex, France}
\altaffiltext{7}{School of Physics and Astronomy, University of
Birmingham, B15 2TT, UK}
\altaffiltext{8}{Stellar Astrophysics Centre, Department of Physics
and Astronomy, Aarhus University, Ny Munkegade 120, DK-8000 Aarhus C,
Denmark}
\altaffiltext{9}{Laboratoire AIM, CEA/DSM, CNRS, Universit\'{e}
Paris Diderot, IRFU/SAp, Centre de Saclay, 91191 Gif-sur-Yvette
Cedex, France}
\altaffiltext{10}{Space Science Institute, Boulder, CO 80301, USA}

\begin{abstract}
\SCAB\ are among the brightest stars observed by {\it Kepler}. What makes these 
stars more interesting is that they are solar analogs. \SCAB\ exhibit solar-like 
oscillations. In this work we use oscillation frequencies obtained using 2.5 
years of {\it Kepler} data to determine the current helium abundance of these 
stars. For this we use the fact that the helium ionization zone leaves a 
signature on the oscillation frequencies and that this signature can be calibrated 
to determine the helium abundance of that layer. By calibrating the signature 
of the helium ionization zone against models of known helium abundance, 
the helium abundance in the envelope of \SCA\ is found to lie in the range 
0.231 to 0.251 and that of \SCB\ lies in the range 0.218 to 0.266.
\end{abstract}


\section{Introduction}
\label{sec:intro}

Helium is the second most abundant element in normal stars and  has a pronounced 
effect on their structure and evolution. Unfortunately, helium abundance cannot 
be determined spectroscopically for low-mass stars because of their low 
temperatures. As a result, the initial helium abundance of stellar models 
depends on an assumption about galactic chemical evolution. For instance, 
the Yale-Yonsei Isochrones \citep{dema04} were constructed assuming 
$Y_0=0.23 + 2Z_0$, $Y_0$ and $Z_0$ being respectively the initial helium and 
heavy-element abundances. The Dartmouth tracks \citep{dott08} assume 
$Y_0 = 0.245+1.54 Z_0$, while the Padova tracks of \citet{mari08} assume 
$Y_0= 0.23+2.23Z_0$. Such ad hoc prescriptions could easily lead to large errors 
in results for single stars that are determined using such sets of isochrones and
tracks.

When additional data, such as frequencies of stellar oscillations and other 
spectroscopically determined stellar parameters are available, the helium abundance 
is often treated as one of the free parameters that are adjusted to obtain the 
best fit to the data \citep[see e.g.,][etc]{metc10,metc12,math12,grub13}. While 
such methods give better constraints on the initial helium abundance, there are 
possible systematic errors that are introduced because of our inability to model 
the near surface layers of a star properly. Normal one-dimensional models treat convection 
only in gross approximations and do not treat the dynamical effects of convection 
at all. This leads to large differences in the structure of the near surface 
layers that introduces a frequency-dependent error in the frequencies 
\citep[see, e.g.,][]{jcd91} that is usually referred to as the ``surface effect''.
Removal of the surface effect can introduce errors in the determined stellar 
properties.

There are however, more direct ways of determining the helium abundance of a star
using its oscillation frequencies. The process of ionization of helium locally 
depresses the adiabatic index, $\Gamma_1$, which in turn affects the sound speed 
in that region, and consequently, the frequencies of acoustic modes. In the case 
of the Sun, the signature that helium ionization leaves on the sound-speed profile 
has been used successfully to determine the helium abundance in the solar 
convection zone \citep[e.g.,][and references therein]{basu95}. For the Sun, the 
helioseismic estimation of helium abundance was facilitated by the fact that the 
frequencies of about a few thousand modes of degrees ranging from $l=0$ (the 
radial mode) to $l=250$ have been determined precisely. This is not the case for 
other stars, where only modes of $l=0$--2 and sometimes $l=3$ can be determined.
Thus the estimation of helium abundance has to be done with a different technique.

To estimate the helium abundance in other stars, one can exploit the fact that the 
steep variation in the sound speed in the second helium ionization 
zone introduces an oscillatory component, $\delta \nu$, in the oscillation frequencies 
as a function of the radial order, $n$, of the modes \citep{goug88,voro88,goug90}.
The signal varies as $\sin(4\pi\tau_\mathrm{He}\nu+\phi)$, where $\tau_\mathrm{He}$ 
is the acoustic depth of the ionization layer as measured from the stellar surface, i.e.,
\begin{equation}
\tau_\mathrm{He}=\int_{r_\mathrm{He}}^{R_*} {dr\over c}\;,
\label{eq:tau}
\end{equation}
$c$ is the sound speed, $r_\mathrm{He}$ the radial distance to the ionization layer,
$R_*$ is the stellar radius, $\nu=\nu_{n,l}$ is the frequency of a mode with
radial order $n$ and degree $l$, and $\phi$ is the phase of the oscillatory term.
The amplitude of the signal is predominantly a function of the helium abundance 
--- higher the abundance, higher the amplitude; but there is also an effect of 
mass, effective temperature, and luminosity, which is why it is important to 
choose close enough calibrating models. \citet{basu04}, \citet{houd04}, and 
\citet{mont05} explored theoretically how this signature may be used to determine 
the helium abundance of other stars. While thus far no one has used this signature 
to determine the helium abundance of any star other than the Sun, it has been
used to determine the acoustic depth of the helium ionization zone. Using data from 
{\it CoRoT}, \citet{migl10} determined the location of the second helium ionization 
zone for the red giant HR~7349, they lacked sufficient signal to noise to determine 
the helium abundance. Again using {\it CoRoT} data, \citet{roxb11} and 
\citet{mazu11,mazu12} determined the position of the HeII ionization zone of the 
solar-type star HD~49933. \citet{mazu14} have extended this study to 19 stars 
observed by {\it Kepler}. Note that a similar oscillatory signature, but at a 
larger acoustic depth $\tau_\mathrm{CZ}$ arises from the discontinuity in the 
second derivative of the sound speed at the base of the convection zone and can be 
used to determine the depth of the convection zone \citep{ball04,mazu05,piau05}. 

In this work we determine the helium abundance in \SCAB\ (HD~186408 and 186427; 
KIC~12069424 and 12069449) using oscillation frequencies determined from 2.5-years 
of observations by the {\it Kepler} satellite. These stars form a binary system 
and are among the brightest stars in the {\it Kepler} field of view. Analysis of 
a three month time series to constrain the properties of the stars have shown 
that the stars are slightly more massive than the Sun and somewhat older too. 
Using the frequencies of oscillations, along with spectroscopic constraints on the 
effective temperature and metallicity of the stars, \citet{metc12} found the stars 
to be $6.8\pm 0.4$ Gyr old  with an initial composition to be $Z_0=0.024\pm0.002$ 
and $Y_0=0.25\pm 0.01$. They estimated the masses of the A and B components to be 
$1.11\pm0.02$ \msun\ and $1.07\pm0.02$ \msun, respectively. Using similar technique 
\citet{grub13} found a somewhat larger helium abundances for these stars. However, 
these values of helium abundance depend on input physics of stellar models and on how the 
surface effects are corrected for. A more direct determination of the current 
helium abundance in the envelope, $Y$, is thus in order.  

The rest of the paper is organized as follows. We describe the methods we have 
used to determine the helium abundance from the signature of the ionization zone 
in Section~\ref{sec:tech}, the models used for the calibration of the helium 
signature are described in Section~\ref{sec:mod}, the results are presented in 
Section~\ref{sec:res} and we discuss the results in Section~\ref{sec:con}.

\section{Extracting the helium signal from the data} 
\label{sec:tech}

The amplitude of the oscillatory term, introduced by the localized depression of 
$\Gamma_1$ in the helium ionization zone, is quite small, and different techniques 
have been used to extract this signal. In this work we use three techniques, 
the first (Method A) relies on taking the second differences of the frequencies 
of a given degree with respect to the  radial order $n$ to suppress the smooth 
variation in frequencies. The other two methods (Methods B and C) rely on directly 
fitting the oscillatory signature while simultaneously modelling the predominantly 
smooth behavior of the frequencies as a function of $n$. The difference between 
the Methods B and C is in the form of oscillatory function used in fitting and 
how the smooth component is adjusted. We describe these methods in more detail 
below.

\subsection{Technique using second differences (Method A)}
\label{subsec:techa}

To extract the oscillatory signal from the frequencies we take the second difference 
\citep{goug90,basu94,basu04} of the frequencies with respect to the radial order 
$n$, i.e.,
\begin{equation}
\delta^2\nu_{n,l}=\nu_{n-1,l}-2\nu_{n,l}+\nu_{n+1,l}.
\label{eq:dif}
\end{equation}
The advantage of taking the second difference is that the contribution from the 
smooth variation of the frequencies with $n$ is greatly reduced since the 
dominant variation of $\nu$ with $n$ is linear.

We fit the second differences to a suitable functional form that represents the
oscillatory signal from the HeII ionization zone as well as that from the base of 
the convection zone \citep{mazu01}. For this purpose we use the following form 
which has been adapted from \citet{houd07}
\begin{eqnarray}
\delta^2\nu&=& a_0+a_1\nu+{b_2\over\nu^2}\sin(4\pi\nu\tau_\mathrm{CZ}+\psi_\mathrm{CZ})\nonumber\\
&&\qquad + c_0\nu \; e^{-c_2\nu^2}\, \sin(4\pi\nu\tau_\mathrm{He}+\psi_\mathrm{He}),
\label{eq:fit_eq_a}
\end{eqnarray}
where $a_0, a_1, b_2, c_0, c_2, \tau_\mathrm{CZ}, \tau_\mathrm{He}, \psi_\mathrm{CZ}, 
\psi_\mathrm{He}$ are free parameters which are fit to the observed second 
differences. In Eq.~(\ref{eq:fit_eq_a}) the first two terms represent the smooth 
part of the function, which remains after the second differences are calculated 
and includes a contribution from the hydrogen and first helium ionization zone; 
the third term represents the contribution from the base of the convection zone, 
while the last term represents the contribution from the HeII ionization zone. 
It is possible to add more terms to the smooth part using a higher degree polynomial,
but for both stars considered in this work, the reduction in
$\chi^2$ due to additional terms was found to be statistically insignificant;
the linear term was found to have a significant effect. Hence, we have
restricted the smooth part to linear form in this work.

The parameters of Eq.~(\ref{eq:fit_eq_a}) are determined by a nonlinear least-squares 
fit. Since the second differences of a given value of $l$ and neighboring $n$ 
have correlated errors, we use the full covariance matrix to calculate the value 
of $\chi^2$. We assume that the errors in the frequencies themselves are not 
correlated with each other to obtain the covariance matrix. Since the nonlinear 
minimization does not always converge to the global minimum of the parameter 
space, we repeat the minimization 100 times with different initial guesses which 
are randomly selected in some range of possible values. The minimum of the $\chi^2$ 
values obtained for the  100 trials is accepted as the best fit value.

To estimate the uncertainties in the fitted parameters we repeat the whole process 
for 1000 realizations of the data obtained by adding random perturbations to the 
frequencies with standard deviation equal to $1\sigma$ uncertainty in the 
corresponding frequencies. The median value of each parameter for 1000 realizations 
was accepted as the fitted value, while the $\pm1\sigma$ uncertainty in the fitted 
value was estimated from the range covering 34\% area on either side of 
the median in the distribution function of the fitted parameter. For the two stars 
being analyzed we find that while the signal from HeII ionization zone is robust 
and all realizations of the data result in values of $\tau_\mathrm{He}$ in a 
reasonable range, the distribution of the fitted values of $\tau_\mathrm{CZ}$ 
is extremely wide with multiple peaks. We find that the oscillatory signal from 
the base of the convection zone is weak for both \SCAB. For some of the 
realizations we find that fitting $\tau_\mathrm{CZ}$ is difficult. An analysis 
of the peaks reveals that one of the secondary peaks corresponds to
$\tau=T_0-\tau_\mathrm{CZ}$ where $T_0$ is the acoustic radius of the star. This 
can happen due to aliasing \citep{mazu01}. We discard all realizations for which 
the fitted value of $\tau_\mathrm{CZ}$ does not lie in the dominant peak of the 
distribution of $\tau_\mathrm{CZ}$. The median of the dominant peak is used as 
the estimate for $\tau_\mathrm{CZ}$. For determining other fitted parameters too we 
use only those realizations that have $\tau_\mathrm{CZ}$ in the dominant peak.
Note that this paper is about determining the helium abundance of these stars, 
and not the position of the convection-zone base, though that is by itself an 
interesting question. To this effect we have tested whether our inability to 
fit $\tau_\mathrm{CZ}$ reliably affects the helium results by fixing the value 
of $\tau_\mathrm{CZ}$ at different values around the main peak. The fit to the 
HeII signal is unaffected, and therefore we are certain that our results 
concerning helium are not biased.

The amplitude of the helium signal in the second differences can be converted to 
that in the frequencies by dividing the second-difference amplitude by a factor
$4\sin^2(2\pi\tau_\mathrm{He}\Delta_0)$, where $\tau_\mathrm{He}$ and $\Delta_0$ 
are the acoustic depth and the average large frequency separation respectively.
This factor is derived under the assumption that the amplitude of oscillatory 
signal varies slowly over the `wavelength' of oscillation.
The amplitude of the oscillatory signal depends on the helium abundance in the 
ionization zone which in stars that have an outer convection zone is the helium 
abundance of the convective envelope. As shown by \citet{basu04}, the amplitude 
of the signal can be calibrated using models with known helium abundance. 
\citet{basu04} used the average amplitude over the frequency interval that is 
used in fitting the signal to measure the helium abundance; we do the same. 

\subsection{Technique using frequencies directly (Method B)}
\label{subsec:techb}

To extract directly the signal from the oscillation frequencies, we adapt a method 
first proposed by \citet{mont94}. The approach is, first, to fit a smooth function 
of radial order $n$ to all the modes $\nu_{n,l}$ of same degree $l$, which removes 
the slowly varying trend. For each degree $l$, we fit a polynomial 
$P_{l,\mathrm{B}}(n)$ using least-squares method with third derivative smoothing. 
The resulting smooth function removes mainly variation of $\nu_{n,l}$ at much 
longer scales than the characteristic scales of the oscillatory signals, filtering 
out the signatures of the acoustic glitches. The filtered signal is then fitted 
simultaneously for all degrees to the following expression:
\begin{eqnarray} 
\delta\nu &=& A_\mathrm{CZ} \left( \frac{\nu_r}{\nu} \right)^2
    \cos \big( 4\pi\tau_\mathrm{CZ} \nu + \phi_\mathrm{CZ} \big) \nonumber \\
  && + A_\mathrm{He} \left( \frac{\nu_r}{\nu} \right)
    \sin^2(2\pi\beta_\mathrm{He} \nu) 
    \cos \big(4\pi\tau_\mathrm{He} \nu + \phi_\mathrm{He}\big),
\label{eq:fit_eq_b}
\end{eqnarray}
where $A_\mathrm{CZ}$, $\tau_\mathrm{CZ}$, $\phi_\mathrm{CZ}$, $A_\mathrm{He}$, 
$\beta_\mathrm{He}$, $\tau_\mathrm{He}$, $\phi_\mathrm{He}$ are 7 free parameters, 
while $\nu_r$ is a reference frequency. 

We have used a modified version of the method used in previous works 
\citep{mont94,mont98,mont00}. It differs mainly in three aspects: (1) a global 
minimization method, based on the PIKAIA implementation of a genetic algorithm 
\citep{char95}, is used for the least-squares fit providing independence of 
initial conditions; (2) the robustness of the fit against outliers is improved 
by performing an iteratively reweighted least-squares regression 
\citep[e.g.,][]{stre88}; and (3) both signals from the base of the convection 
zone and the HeII ionization zone are fitted simultaneously.

The third derivative smoothing procedure depends on a parameter $\lambda_\mathrm{B}$ 
which determines how closely the polynomial $P_{l,\mathrm{B}}(n)$ interpolates 
the points \citep[their Eq.~C5]{mont94}. Here, we no longer iterate on this 
parameter but set it to a value adequate for isolating the longest period signal 
from the HeII ionization zone. The same value of $\lambda_\mathrm{B}$ is used for 
both stars, either observations or model frequencies. With this method, all 
available data points are used (frequencies of consecutive orders are not required 
to construct combinations) and we do not \emph{a priori} assume a specific 
functional form for the smooth component. 

To estimate the uncertainties in the fitted parameters we perform 5000 Monte 
Carlo realizations of the frequencies by adding random perturbations to them with 
standard deviation equal to $1\sigma$ uncertainty in the corresponding frequencies. 
The median value of the distribution of each parameter was accepted as the fitted 
value and the $\pm 1 \sigma$ uncertainty in that parameter was estimated from the 
standard deviation of the distribution.

To calibrate the helium abundance, we used the amplitude of the signal at a 
reference frequency $\nu_r$ ($=2.0$~mHz) for both components. This choice provides 
us a simple measure of the amplitude, $A_\mathrm{He} \sin^2(2\pi\beta_\mathrm{He} \nu_r)$ 
(see Eq.~\ref{eq:fit_eq_b}), and hence a measure of the amount of helium present 
in the ionization zone. 

\subsection{Technique using frequencies directly (Method C)}
\label{subsec:techc}

This technique is a modification of the previous one in the sense that it fits 
smooth and oscillatory signal together. For each $l$, we take a fourth degree 
polynomial in $n$, $P_{l,\mathrm{C}}(n)$, for the smooth component. Although, a 
third degree polynomial takes care of smooth component for the observed data, 
it is not enough for the models, as the surface term introduces a smooth component 
which varies approximately as fourth power of the frequency \citep{kjel08}. We 
fit directly the frequencies to the function
\begin{eqnarray}
f(n,l) &=& P_{l,\mathrm{C}}(n) 
              + \frac{A_\mathrm{CZ}}{\nu^2} \sin(4\pi\tau_\mathrm{CZ}\nu + \phi_\mathrm{CZ}) \nonumber \\
              && + A_\mathrm{He} \nu e^{-c_{2}\nu^2} \sin(4\pi\tau_\mathrm{He}\nu + \phi_\mathrm{He}) \,,
\label{eq:fit_eq_c}
\end{eqnarray} 
where $P_{l,\mathrm{C}}(n) = \sum_{i = 0}^{4} A_{l,i} n^{i}$. The $4\times5$ 
elements of $[A]_{l,i}$ along with $A_\mathrm{CZ}$, $\tau_\mathrm{CZ}$, 
$\phi_\mathrm{CZ}$, $A_\mathrm{He}$, $c_2$, $\tau_\mathrm{He}$, $\phi_\mathrm{He}$ 
are 27 free  parameters which need to be determined.

The parameters of Eq.~(\ref{eq:fit_eq_c}) are determined by minimizing a $\chi^2$ 
defined as:
\begin{equation}
\chi^2 = \sum_{n,l}\left[\frac{\nu_{n,l}-f(n,l)}{\sigma_{n,l}}\right]^2 + 
\lambda_\mathrm{C}^2 \sum_{n,l}\left[\frac{d^2P_{l,\mathrm{C}}(n)}{dn^2}\right]^2\,,
\end{equation}
where $\sigma_{n,l}$ is the uncertainty in the mode frequency and $\lambda_\mathrm{C}$
is a smoothing parameter. A very small value of $\lambda_\mathrm{C}$ corresponds to the 
standard weighted least-squares fitting, while a very large value tends to fit the 
frequencies to a straight line. To make an appropriate choice of $\lambda_\mathrm{C}$, 
we note that as $\lambda_\mathrm{C}$ increases, the fit to the observation stabilizes 
(uncertainty in the fitting parameters decreases) and saturates at some point. We 
accept a value of $\lambda_\mathrm{C}$ when the stabilization occurs. The value of 
$\lambda_\mathrm{C}$ is found to be the same for both stars. To avoid systematic 
errors in the calibration, we use the same $\lambda_\mathrm{C}$ for the models as well. 
While the number of parameters to be fitted in this method is large,
it should be noted that the use of regularization introduces additional
constraints that enable more parameters to be determined.

Similar to what has been done in Method A, we repeat the minimization 100 times 
with random initial guesses covering a reasonable parameter space to get the 
global minimum. We simulate 1000 realizations of the data in the same way as was 
done in Method A to get the fitted value along with $\pm 1\sigma$ uncertainty. 
We calculate an average amplitude over the frequency range used in the fitting 
for the calibration.

\section{Stellar models for calibration}
\label{sec:mod}

To determine the helium abundance of the stars we need to compare the observed 
amplitudes of the helium signal to that calculated for stellar models with different 
helium abundances. We used three sets of models: two constructed with the MESA 
code \citep{paxt11} for two different metallicity mixture, and one with the Yale 
stellar evolution code \citep[YREC;][]{dema08}. The models in each set were 
constructed following different principles and the models for calibration were 
selected using different criteria. The models and the selection process are 
described below.

\subsection{The MESA models}
\label{subsec:mesa}

The MESA code allows users to explore different input physics and data. Our models
were constructed using OPAL equation of state \citep{roge02}. We used OP opacities 
\citep{badn05,seat05} for models with metallicity mixture of \citet{gs98} 
(hereinafter GS98) and OPAL opacities \citep{igle96} for models with metallicity mixture of 
\citet{aspl09} (hereinafter AGSS09). Low temperature opacities were used from 
\citet{ferg05} assuming GS98 mixture. Reaction rates from NACRE \citep{angu99} 
were used for all reactions except $^{14}$N($p$,$\gamma$)$^{15}$O and 
$^{12}$C($\alpha$,$\gamma$)$^{16}$O, for which updated rates of reaction from 
\citet{imbr05} and \citet{kunz02} were used. Convection was modelled using the 
standard mixing length theory \citep{cox68}, and diffusion of helium and heavy 
elements was incorporated using the prescription of \citet{thou94}.

We constructed models assuming two different values of the solar metallicity to 
convert the observed \febyh\ to $Z$ needed for the models. Recall that there is 
currently an uncertainty about what the solar metallicity really is. The solar 
abundances of GS98, with $Z/X=0.023$, had been used extensively for many years 
and solar models constructed with these abundances satisfied helioseismic 
constraints quite well \citep[see e.g.,][]{basu08}. However, \citet{aspl05} redetermined 
the solar abundances using 3D model atmospheres and incorporating NLTE corrections 
for many lines and claimed that the solar abundance of heavy elements was much 
lower, $Z/X=0.0165$. The relative abundances of different elements were also 
changed. \citet{aspl09} revised the abundances further to $Z/X=0.018$. The lower 
abundances are not consistent with helioseismic constraints 
\citep[see e.g.,][]{turc04,bahc05,bahc06,dela06,basu08,jcd09,basu13}
and this discrepancy has not been resolved. To estimate the effect of this 
uncertainty on stellar models we used both GS98 and AGSS09 mixtures of heavy 
elements in this work. Opacities were calculated assuming the different heavy-element 
mixtures, and $Z$ for the models were derived for a given \febyh\ assuming that
$\febyh=0$ corresponds to $Z/X=0.023$ for the GS98 mixture and $Z/X=0.018$ for the 
AGSS09 mixture.

We constructed models on a uniform grid of stellar parameters (mass $M$, initial 
helium abundance $Y_0$, initial metallicity $\febyh_0$, and mixing length parameter 
$\alpha$) for both stars with two different metallicity mixtures, GS98 and AGSS09. 
The range of values in the grid for these parameters were: 
$M: 0.91$\,--\,$1.17\,\msun$,
$Y_0: 0.22$\,--\,$0.35$,
$\febyh_0: 0.12$\,--\,$0.26$,
$\alpha: 1.9$\,--\,$2.3$.
The step sizes were $0.02$\,\msun, $0.02$, $0.02$\,dex and $0.05$, respectively. 
A reduced $\chi^2$ between the frequencies of the model ($\nu_\mathrm{mod}$, 
corrected for the surface term as per the formulation of \citet{kjel08}) and the 
observed frequencies ($\nu_\mathrm{obs}$) is defined as:
\begin{equation}
\chi^2_\nu = \frac{1}{N}\sum\limits_{i} \left(\frac{\nu_\mathrm{obs} -
\nu_\mathrm{mod}}{{\sigma_\mathrm{obs}}}\right)^2_{i}\,,
\label{eq:chisq_nu}
\end{equation}
where $\sigma_\mathrm{obs}$ is the $1\sigma$ uncertainty in the observed frequency
and the sum runs over $N=12$ radial modes of the concerned star. A model with a 
particular $M$, $Y_0$, $Z_0$, and $\alpha$ is selected as an acceptable model of 
the star only if the evolutionary track enters in a 4-dimensional box formed by 
the 2$\sigma$ uncertainties in the observed effective temperature, luminosity, 
surface \febyh, and large separation as given in Table~\ref{tab:obs}. Further 
degeneracy in the age is lifted by minimizing the $\chi^2_\nu$ (Eq.~\ref{eq:chisq_nu}). 
In other words, the ``best'' model on each evolutionary track is chosen through 
a matching of the frequencies, provided the parameters mentioned above agree 
within very conservative limits. The surface metallicity given by \citet{rami09}, 
$0.096\pm0.026$ dex for \SCA\ and $0.052\pm0.021$ dex for \SCB, is not quite 
the same for both components. However, since the metallicities are quite similar, 
we assume the metallicity of the two stars to be equal, $0.096\pm0.040$ dex, with 
increased uncertainty. This may be justified as the stars form a binary system and 
hence their initial metallicities should be the same. Since we use multiple values 
of initial conditions ${M, Y_0, Z_0, \alpha}$, we get two ensembles each, 
corresponding to the two abundance choices of GS98 and AGSS09, for \SCA\ and 
\SCB. These are not, however, the final sets of calibration models. The dispersion 
in the age of the models in each of these ensembles is nearly 5~Gyr. We use a 
further selection procedure described below which takes into account the fact 
that \SCAB\ should have the same age. Models constructed with the AGSS09 mixture 
are treated separately from those constructed with the GS98 mixture so that the 
effect of the uncertainty in the solar metallicity on our results can be quantified.

We constrain the ages using the small frequency separation ($d_{l\,l+2}(n) = \nu_{n,l} - \nu_{n-1,l+2}$), 
which is known to be a good indicator of the evolutionary stage of main sequence 
stars \citep[see e.g.,][]{jcd88}. We use the weighted average of the small 
separation over 12 pairs, $\langle d_{02}\rangle$, for both components. However, we 
do not select the models with $\langle d_{02}\rangle$ alone, but consider other 
observables --- namely, the effective temperature ($T_\mathrm{eff}$), luminosity 
($L$), surface gravity (as $\log g$), surface metallicity (as \febyh), and the 
large separation ($\Delta_0$) --- as well. Although, the radii of the two components 
have been measured with interferometry \citep{whit13}, we do not use the radii 
to constrain the ages, but it should be noted that the interferometric radii are 
found to be consistent with the final model radii for both stars.

We estimate the average large frequency separation ($\Delta_0 = \langle \nu_{n,0} - \nu_{n-1,0} \rangle$) 
of the radial modes by using a linear fit to the frequencies $\nu_{n,0}$ as a 
function of $n$. The same 12 radial modes are used that were used for calculating 
the average small separation. The same set of modes are used to calculate the 
average large separation for the observation and the models. As mentioned earlier, 
the inadequacy of modelling the near-surface layers of the star introduces a 
frequency-dependent shift in the calculated eigenfrequencies, $\nu_{n,l}$ of 
the models, making the frequencies systematically larger than observed frequencies 
for all but the lowest-frequency modes. This error propagates into the large 
separation too. We find that for our acceptable  models $\Delta_0$ is about 
1$\mu$Hz larger than the observed value. We estimated the exact shift using the 
differential form of the Duvall Law, as formulated by \citet{jcd89}, in which the 
frequency differences between a model and observation are decomposed into two 
parts, one that depends on the interior structural differences and another that 
is essentially the surface term. The $\Delta_0$ values of the model are corrected 
for the surface term before comparing them to the observed values of \SCAB.

\begin{figure}
\plotone{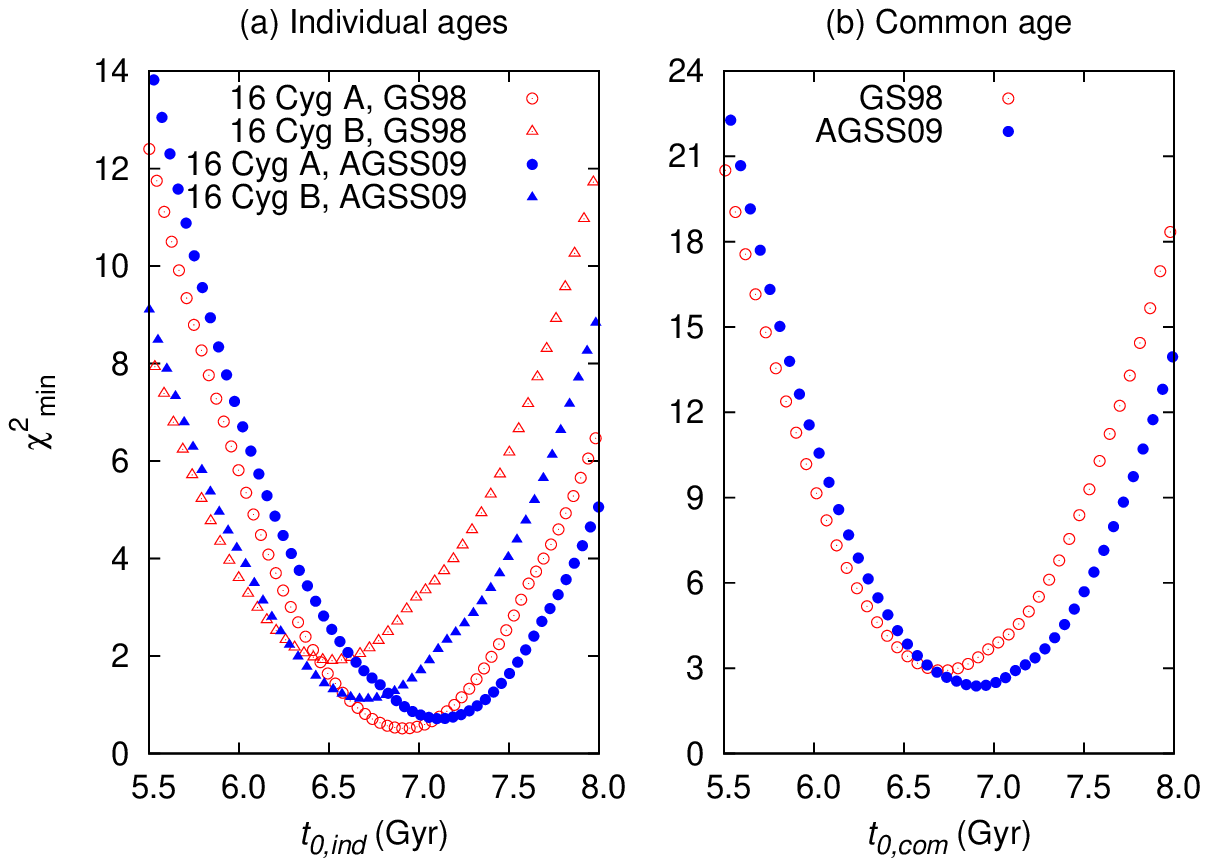} 
\caption{Selecting MESA models with appropriate age: (a) The four different types 
of points correspond to the four model ensembles of \SCAB\ (two for a given star, 
with two metallicity mixtures). A point represents a model that minimizes the 
$\chi_\mathrm{ind}^2$, as defined in Eq.~(\ref{eq:age1}), at a given reference 
age $t_{0,\mathrm{ind}}$. (b) The two different types of points correspond to the 
two model ensembles of \SC\ system with two metallicity mixtures. A point represents 
a pair of models, one from the model ensemble of \SCA\ and the other from \SCB\, 
that minimizes the $\chi_\mathrm{com}^2$, as defined in Eq.~(\ref{eq:age}), at a 
given reference common age $t_{0,\mathrm{com}}$.}
\label{fig:cage}
\end{figure}

We estimated independently the age of \SCA\ and \SCB\ using models of two different 
metallicity mixture. To estimate the age of a star using models of a particular 
metallicity mixture, we define a $\chi^2$ for each model as:
\begin{equation}
\chi_\mathrm{ind}^2 = \sum\limits_{q}\frac{(q_\mathrm{mod}-q_\mathrm{obs})^2}{\sigma_{q}^2} +
\frac{(t_\mathrm{mod}-t_{0,\mathrm{ind}})^2}{\sigma_t^2}.
\label{eq:age1}
\end{equation}
The term $q$ in the summation represents 6 observable quantities ($\Delta_0$, 
$\langle d_{02}\rangle$, $T_\mathrm{eff}$, $L$, $\log g$, and \febyh) and 
$\sigma_q$ denotes the uncertainty in the estimate of the respective quantity. 
The quantity $t_\mathrm{mod}$ is the age of the model under consideration 
and $t_{0,\mathrm{ind}}$ is a reference age. For a given $t_{0,\mathrm{ind}}$, we 
can find a model that represents the star most closely by minimizing the 
$\chi_\mathrm{ind}^2$ (Eq.~\ref{eq:age1}) over all the models in the ensemble. 
We denote the minimum by $\chi_\mathrm{min}^2$, and minimize it with respect 
to $t_{0,\mathrm{ind}}$ in a range spanned by the model ages in the ensemble, 
to get the individual age of the star. For clarity we show the 
$\chi_\mathrm{min}^2$ in Fig.~\ref{fig:cage}(a) in a limited range of 
$t_\mathrm{0,ind}$. Clearly the curves have a well defined minimum that gives 
the best model as well as the best estimate of the age of each individual star 
for a specific assumed metallicity mixture. The ages are respectively $6.9 \pm 0.4$ 
Gyr and $6.6 \pm 0.4$ Gyr for \SCAB\ when GS98 mixture is used. For the AGSS09 
mixture the ages are respectively $7.1 \pm 0.4$ Gyr and $6.7 \pm 0.4$ Gyr. Thus 
the ages of the two stars are found to be close to each other and the values are
consistent with those of \citet{metc12} and \citet{grub13}. Note that the calculation 
of $\chi_\mathrm{ind}^2$ requires $\sigma_t$; we start with a suitable guess and then 
whole process is iterated until we get the same uncertainty in the estimated age as 
$\sigma_t$.

The individual ages of the two stars, independently determined as explained above,
need not turn out to be the same. However, since the stars form a binary system it 
is more natural to assume that their ages are the same. We can determine this common 
age by modifying the definition of $\chi^2$ to include both stars. Thus for a 
given pair of models, with one model drawn from the model ensemble of \SCA\ and 
the other from that of the \SCB, we define a combined $\chi^2$ as:
\begin{eqnarray}
\chi_\mathrm{com}^2&=&
\sum\limits_{q_\mathrm{A}}
\frac{(q_\mathrm{A,mod}-q_\mathrm{A,obs})^2}{\sigma_{q_\mathrm{A}}^2} +
\frac{(t_\mathrm{A,mod}-t_{0,\mathrm{com}})^2}{\sigma_t^2} 
\nonumber\\
&+&
\sum\limits_{q_\mathrm{B}}
\frac{(q_\mathrm{B,mod}-q_\mathrm{B,obs})^2}{\sigma_{q_\mathrm{B}}^2} +
\frac{(t_\mathrm{B,mod}-t_{0,\mathrm{com}})^2}{\sigma_t^2}.
\label{eq:age}
\end{eqnarray}
The first two terms correspond to the model of the A component while the last 
two correspond to the model of the B component. The summations in the first and 
third terms again have 6 terms each corresponding to the same observables as 
described above. The quantities $t_\mathrm{A,mod}$ and $t_\mathrm{B,mod}$ are 
the ages of the models of \SCA\ and \SCB\ under consideration. The quantity 
$t_{0,\mathrm{com}}$ is a reference age as before, but common to both stars. 
For a given $t_{0,\mathrm{com}}$, we can find a pair of models that represents 
the stars most closely by minimizing the $\chi_\mathrm{com}^2$ (Eq.~\ref{eq:age}) 
over all possible pairs of the two ensembles. We denote the minimum by 
$\chi_\mathrm{min}^2$, and minimize it with respect to $t_{0,\mathrm{com}}$ 
in a range spanned by the model ages in the two ensemble, to get the common 
age of the star. For clarity we show the $\chi_\mathrm{min}^2$ in 
Fig.~\ref{fig:cage}(b) in a limited range of $t_\mathrm{0,\mathrm{com}}$. 
Again, the resulting curves have a well defined minimum that gives the 
best pair of models as well as the best estimate of the common age of the 
stars. The individual models of such a pair are essentially the best models 
of \SCAB. The age of the binary system is estimated to be $6.7 \pm 0.3$ Gyr 
for GS98 mixture, and $6.9 \pm 0.3$ Gyr for AGSS09 mixture. Thus we find that 
the mixture of heavy elements does not affect the age significantly.
The difference in the age between models of two mixtures may become significant
when the stellar parameters and the oscillation frequencies are known more precisely.
\citet{houd11} have pointed out that along with $d_{02}$ we can use coefficients of 
higher order asymptotic formula for oscillation frequency to determine the age more 
precisely as these are more sensitive to the sound speed variation in the stellar core. 
However, it is not clear if these terms can be determined more accurately from the 
currently observed frequencies.

At the end of this selection procedure we are left with approximately $450$ models, 
nearly $100$ models for a star with a particular metallicity mixture, with a range 
of mass, chemical composition, and mixing length parameters, and age within $1\sigma$ 
of the value determined above. For the two metallicity mixtures, GS98 and AGSS09, we 
have two such set of calibrating models.

\subsection{The YREC models}
\label{subsec:yrec}

In addition to the models described above, we also used the Yale stellar evolution 
code to model the two stars. The input physics includes the OPAL equation of state (EOS) 
tables of \citet{roge02}, and OPAL high temperature opacities \citep{igle96} 
supplemented with low temperature opacities from \citet{ferg05}. All nuclear 
reaction rates are from \citet{adel98}, except for that of the $^{14}N(p,\gamma)^{15}O$ 
reaction, for which we use the rate of \citet{form04}. All models included 
gravitational settling of helium and heavy elements using the formulation of 
\citet{thou94}. Models were constructed assuming the heavy element mixture of GS98.

The YREC models were constructed with an approach that was quite different from 
the MESA models. The starting point of the modelling was the mass and radius of the
two stars as determined by \citet{metc12}. We did a Monte Carlo around this mass 
and radius, and the spectroscopically determined $T_\mathrm{eff}$ and metallicity
to obtain a set of values for mass, final radius, final $T_\mathrm{eff}$ and final 
metallicity. YREC was used in an iterative mode. In this mode the  final $T_\mathrm{eff}$ 
and radius for a star of a given mass and metallicity are specified. The code 
iterates over the initial helium abundance $Y_0$ in case the mixing length
parameter $\alpha$ is specified, or alternatively over $\alpha$ if $Y_0$ is specified
until a model with the specified $T_\mathrm{eff}$ and radius is found. Note that this is 
similar to the construction of standard solar models, though in the case of the Sun
the iteration is performed simultaneously over both mixing length parameter and $Y_0$ 
(since the solar age is a known constraint). We made the models both ways --- 
specifying $\alpha$ and determining $Y_0$ and specifying $Y_0$ and determining 
$\alpha$. The initial metallicity of the models was kept higher than the observed 
metallicity in order to account for the depletion of metals by diffusion. The Monte 
Carlo exercise assumed that the width of the distribution in mass and radius was
three times the quoted uncertainty in \citet{metc12}. These models, by construction, 
satisfy the temperature constraint. Models of \SCAB\ were constructed independently.

The final selection of models was done as follows. All models with initial helium 
abundance lower than the Big Bang nucleosynthesis value of $Y_p=0.2477$ \citep{peim07} 
were rejected. Also rejected were models with final metallicities that were different 
from the observed metallicity \citep{rami09} by more than 3$\sigma$. We calculated 
the frequencies for the remaining models. The frequencies were used to calculate the 
frequency ratios $r_{02}$, $r_{10}$, and $r_{01}$ \citep{roxb03,silv11}. These ratios 
are insensitive to the surface term and hence do not require any {\it ad hoc} 
corrections to account for near-surface uncertainties in the models. All models 
that matched the observed ratios to within 3$\sigma$ were selected. The common 
age of the \SC\ system was estimated to be $6.6\pm0.3$ Gyr in the same way as 
described in Section~\ref{subsec:mesa}. This resulted in 231 models for \SCA\ and 
95 models for \SCB\ for calibration after applying the age constraint.

\section{Results}
\label{sec:res}

We applied all three techniques described in Section~\ref{sec:tech} to estimate 
the helium abundance of both components of the \SC\ system. We also estimated 
the acoustic depths of the HeII ionization zone and made preliminary estimates 
of the acoustic depths of the base of the convection zone.

In this work, we used observed frequencies computed following the procedure 
described by \citet{appo12} from 2.5~years of {\it Kepler} simple-aperture-photometry 
light curves corrected following \citet{garc11} and high-pass filtered with a 4-days 
triangular smooth. The observed frequencies as used in this work are given in 
Table~\ref{tab:freq}. All frequencies listed in the table were used in the fits
except for the last two modes for \SCA\ which were not used in Methods A,C as
these modes are beyond the cut-off frequency in many stellar models and hence
cannot be determined reliably in stellar models which are used for calibration.
The Method B is not expected to be
sensitive to presence of these modes in the observed frequency set as the
amplitude of helium signal is compared at a reference frequency instead of
the average over entire range covered in the fit, which is the case in Methods A,C.
Further, the two isolated $l=3$ modes cannot be used in Method A, which needs to compute
second order differences.

\begin{figure}
\plotone{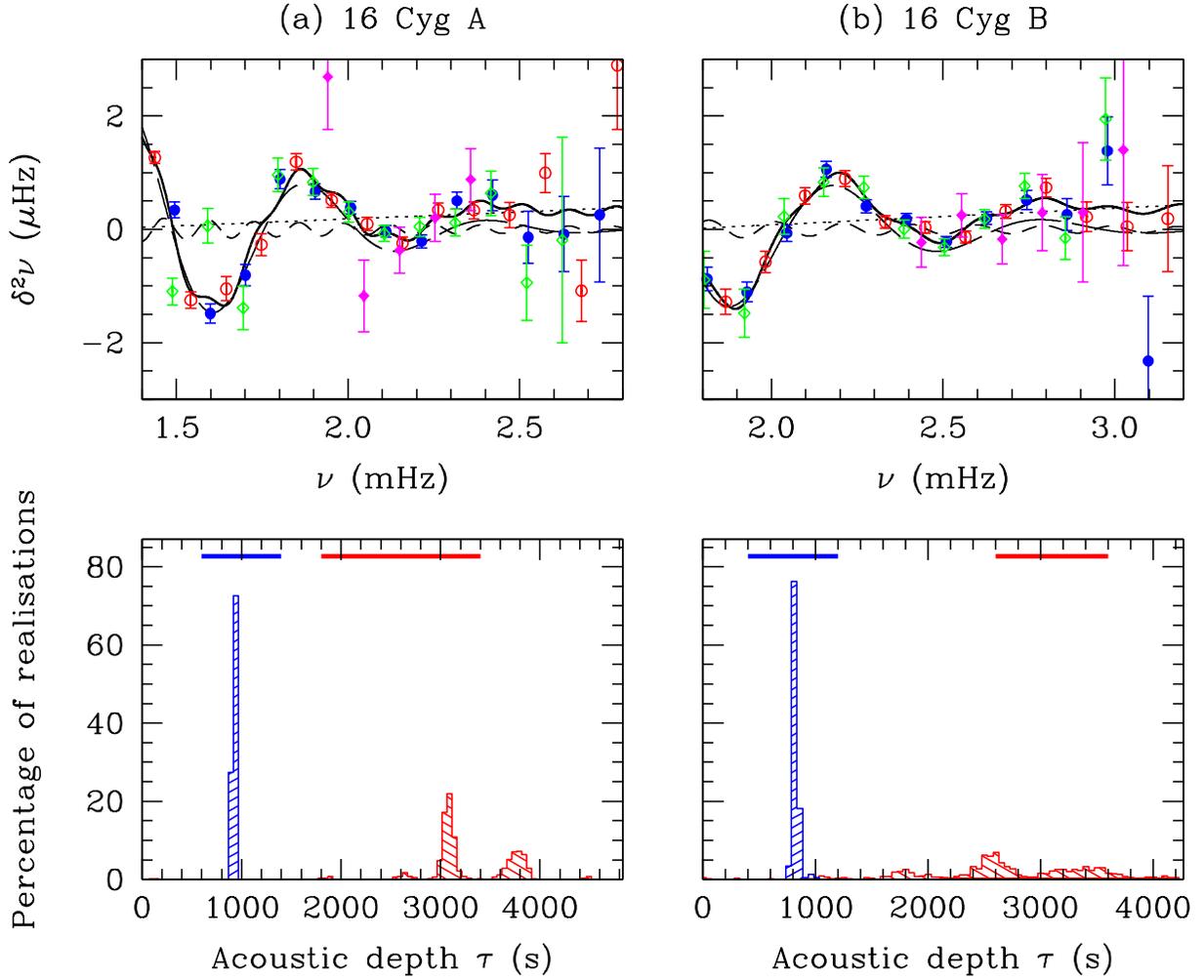}
\caption{Fit to the second differences (Method A) for \SCAB. The upper panels show 
the fits (solid line) as well as the three components on the RHS of Eq.~(\ref{eq:fit_eq_a}). 
The dotted line shows the smooth part, the short-dashed line shows the contribution 
from CZ signal and the long-dashed line shows the contribution from the HeII 
ionization zone signal. The blue filled circles are the second differences for 
$l=0$ modes, the red open circles are for $l=1$, the green open diamonds are for 
$l=2$, and the magenta filled diamonds are for $l=3$. The lower panels show the 
histograms of distribution of $\tau_\mathrm{He}$ (blue) and $\tau_\mathrm{CZ}$ (red).
The horizontal bars at the top show the range of initial guesses for the two 
parameters used for fitting.}
\label{fig:fit_fig_a}
\end{figure}

\begin{figure}
\plotone{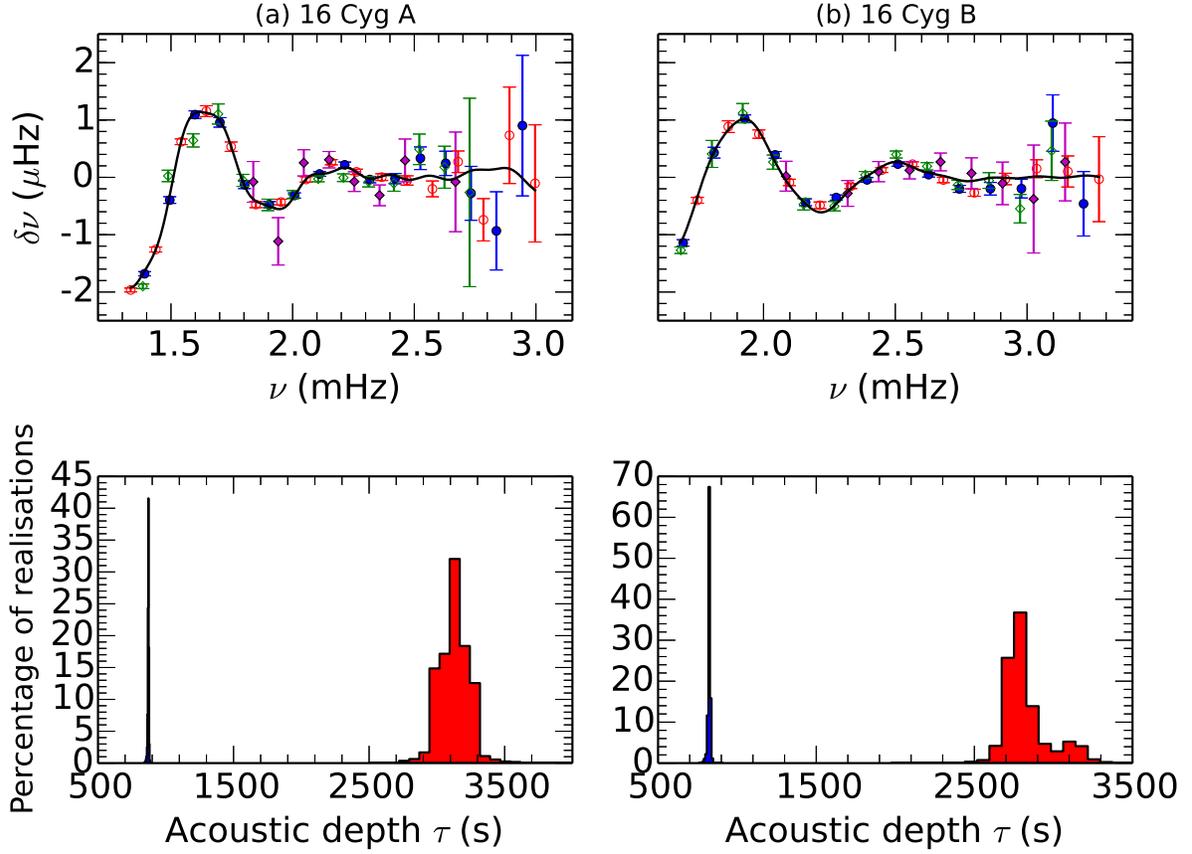}
\caption{Fit to the frequencies (Method B) for \SCAB. The upper panels show the fit 
(solid line) to Eq.~(\ref{eq:fit_eq_b}). The symbols are same as in 
Fig.~\ref{fig:fit_fig_a}. The lower panels show the histograms of distribution of 
$\tau_\mathrm{He}$ (blue) and $\tau_\mathrm{CZ}$ (red).}
\label{fig:fit_fig_b}
\end{figure}

\begin{figure}
\plotone{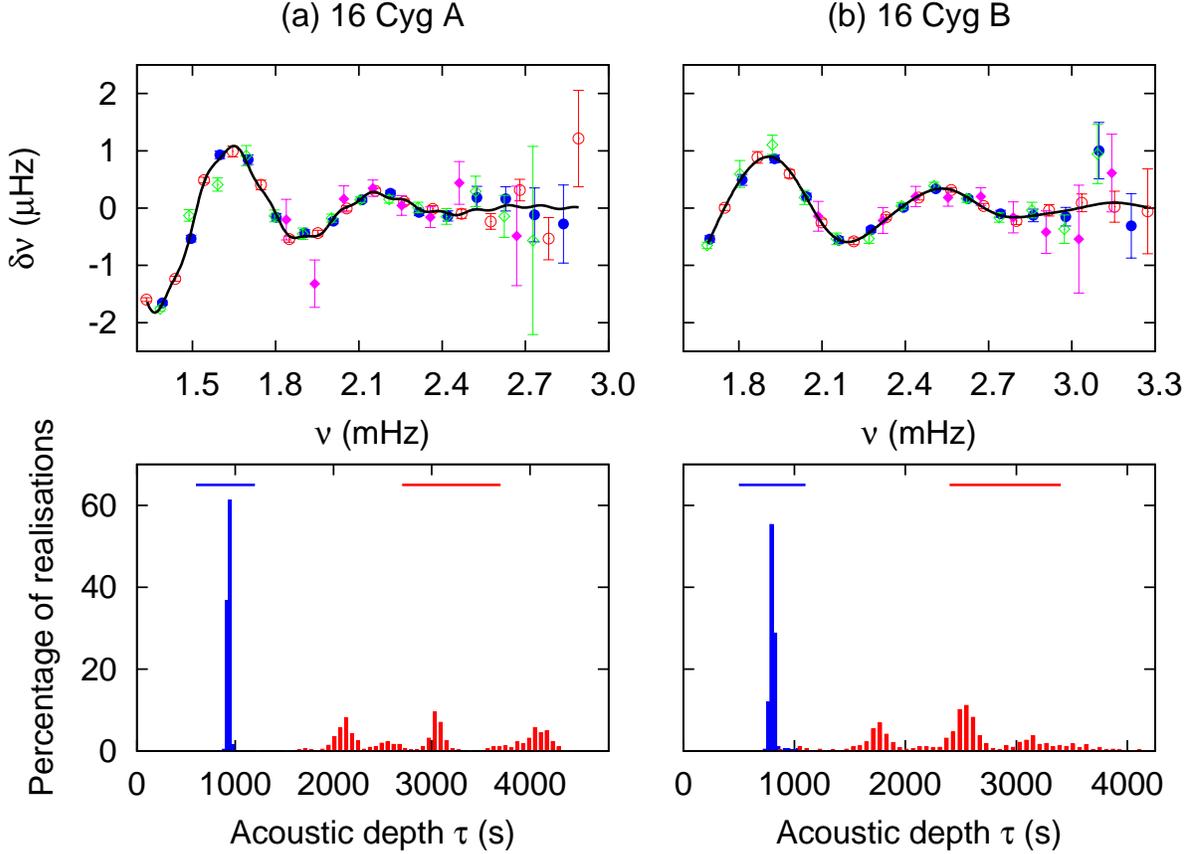}
\caption{Fit to the frequencies (Method C) for \SCAB. The upper panels show the combined 
oscillatory signal obtained by removing the contribution of the polynomial 
$P_{l,\mathrm{C}}(n)$ from the frequencies. The symbols are same as in 
Fig.~\ref{fig:fit_fig_a}. The curve shows $f(n,l)$ without the contribution of 
$P_{l,\mathrm{C}}(n)$. The lower panels show the histograms of distribution of 
$\tau_\mathrm{He}$ (blue) and $\tau_\mathrm{CZ}$ (red). The horizontal bars at 
the top show the range of initial guesses for the two parameters used for fitting.}
\label{fig:fit_fig_c}
\end{figure}

The upper panels of Fig.~\ref{fig:fit_fig_a}, \ref{fig:fit_fig_b}, and 
\ref{fig:fit_fig_c} show respectively the fit to Eq.~(\ref{eq:fit_eq_a}) (Method A), 
Eq.~(\ref{eq:fit_eq_b}) (Method B) and Eq.~(\ref{eq:fit_eq_c}) (Method C) of \SCAB; 
whereas the lower panels show the histograms of distribution of $\tau_\mathrm{He}$ 
(blue) and $\tau_\mathrm{CZ}$ (red) obtained by fitting 1000 realizations (5000 
for Method B) of data perturbed by adding random errors. The fitted values of 
the various physical parameters are listed in Table~\ref{tab:fitpar}. It should 
be noted that for Methods A and C, the amplitude is averaged over the fitting 
interval, while for Method B, it is calculated at a reference frequency.
As explained in Section~\ref{subsec:techa}, the amplitude of oscillatory term in
second differences is converted to that in frequencies. From the table it
can be seen that the amplitudes of oscillatory signals obtained with methods
A and C are similar. This demonstrates the validity of conversion
factor. The fitted parameters using all three methods are close to
each other and generally within $1\sigma$, except for $\tau_\mathrm{He}$ using
method B. Even in this case the difference is only about 6\% of the acoustic depth.
This demonstrates that the results are not particularly sensitive to different
form of oscillatory term or to different techniques for removing the smooth
component in the frequencies as a function of $n$.

\begin{figure}
\plotone{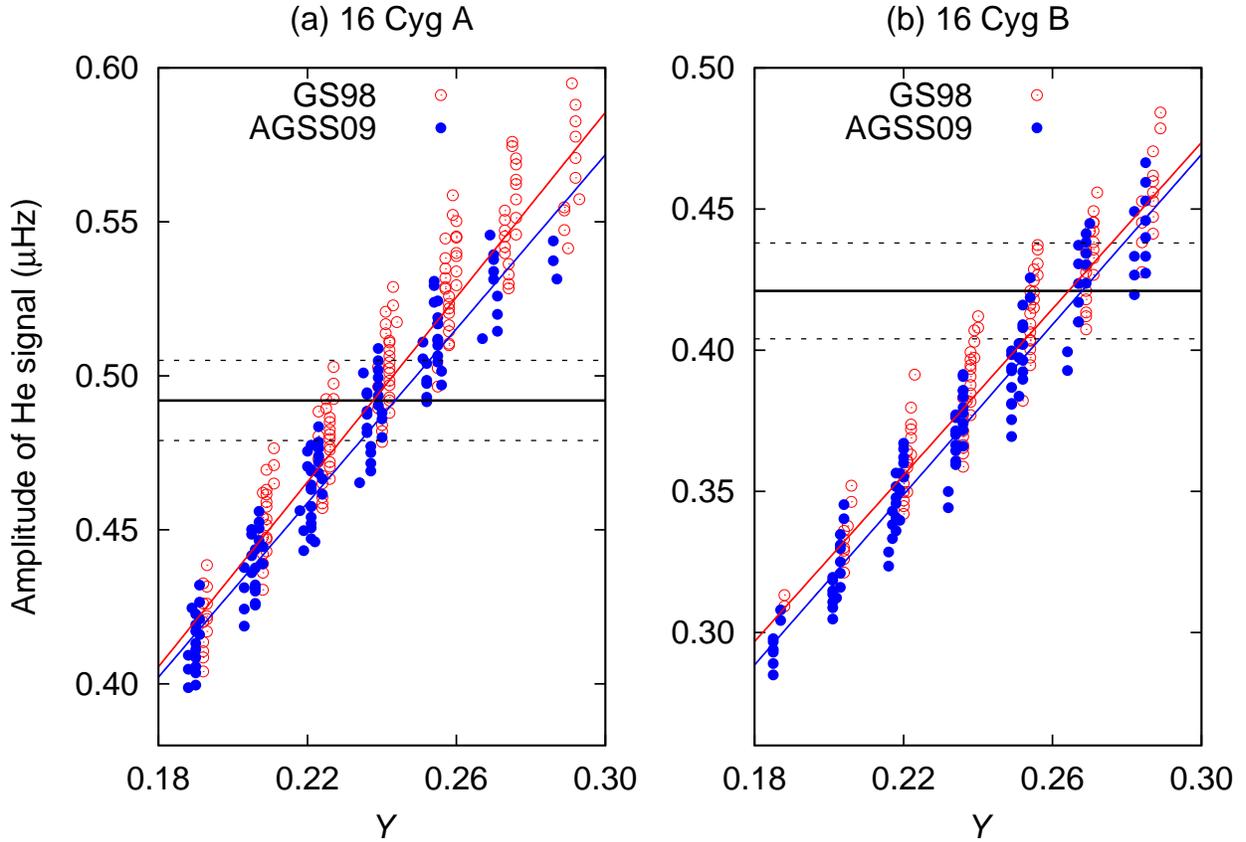}
\caption{Amplitude of the helium signal obtained using Method A as a function of 
$Y$ for \SCAB. The continuous horizontal line is the observed amplitude, with 
the dashed lines representing the $1\sigma$ uncertainty limits. The points are 
the results for the MESA models, while the red and blue lines show the straight 
line fit to these points.}
\label{fig:ampa}
\end{figure}

The frequencies for the models were fitted in the same manner, and using the same 
modes and weights as the observations. Amplitude of the HeII signal obtained using 
Method A is shown in Fig.~\ref{fig:ampa} for each MESA model. Also shown in the 
figure is the observed amplitude. Note from the figure that the amplitude is 
predominantly a function of the current helium abundance in the models. There 
is however, some scatter due to other model parameters, primarily due to variations 
in mass, effective temperature, and luminosity. The helium abundance, $Y$, obtained by 
calibrating the model amplitude with observation are listed in Table~\ref{tab:yresults} 
for both stars. The helium abundance obtained with MESA and YREC are consistent 
with each other within errors.  There is a difference, barely within errors, 
between the helium abundance of \SCA\ and \SCB. For \SCA\ all 3 methods give 
results that are within error bars of each other, but for \SCB\, Method A tends 
to give results that are on higher side, while Method B tends to give lower 
values. The difference between the two methods is up to $2.5\sigma$. The difference 
in results between various methods can be considered to be a measure of systematic 
errors.
The estimate of $Y$ doesn't directly depend on $\tau_\mathrm{He}$ and hence
is not affected by uncertainties in it.

The precision of determination of helium abundance depends critically on the modes 
at the low end of the spectrum. Addition of a few low order modes or any 
significant improvement in the precision of these modes, 
improve the precision of estimate of helium abundance significantly. To understand 
this improvement we repeated the fit using Method C 
after removing successively the lowest order modes, or after increasing
the uncertainties in the lowest three modes of degree 0, 1, and 2. We found that 
the uncertainty in the amplitude of the helium signal increases rapidly in each 
case, by a factor of two or more.
This may be due to the fact that the amplitude of the signal decreases very 
rapidly with increasing frequency, and the low order modes play a role in stabilizing 
the fit as in that region the amplitude of helium signal is larger.
Therefore a precise determination of the helium abundance of a star 
would particularly require a precise set of low order modes.

In Method B, the amplitude of the oscillatory signal at a reference 
frequency (instead of the mean amplitude used in other methods)
is used to calibrate the helium abundance. The stability of this approach
has been tested by \citet{mont05} for the case of the Sun. They found that
the use of a reference frequency well inside the frequency
range of the observations was a better, more stable option, to calibrate $Y$.
This is so because the approach used by this method to account for the smooth
component is fairly robust in the inner region of the frequency range (does not change
regardless of the details for smooth component that is adjusted). In this way, this method
can avoid any contamination from the boundaries due to the smoothing.

\begin{figure}
\epsscale{0.5}
\plotone{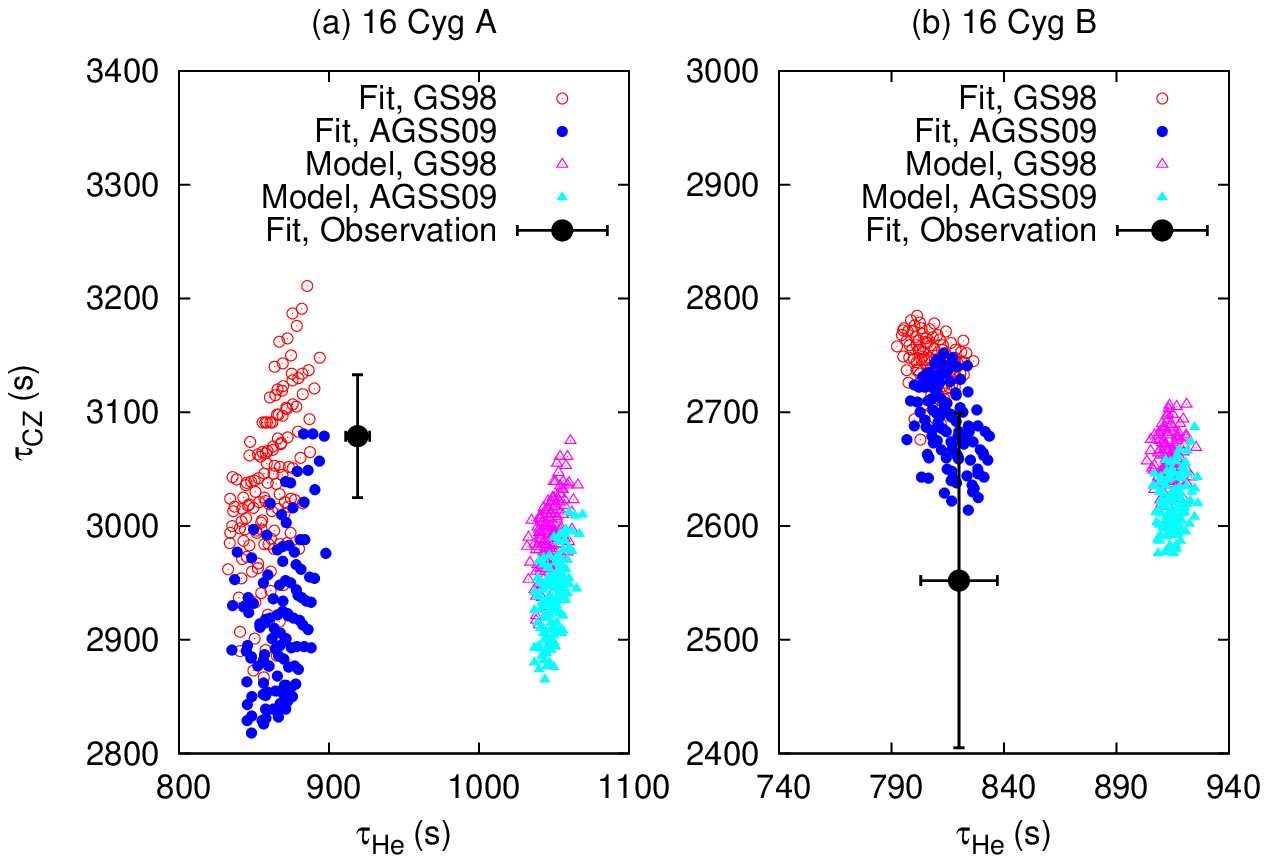}
\plotone{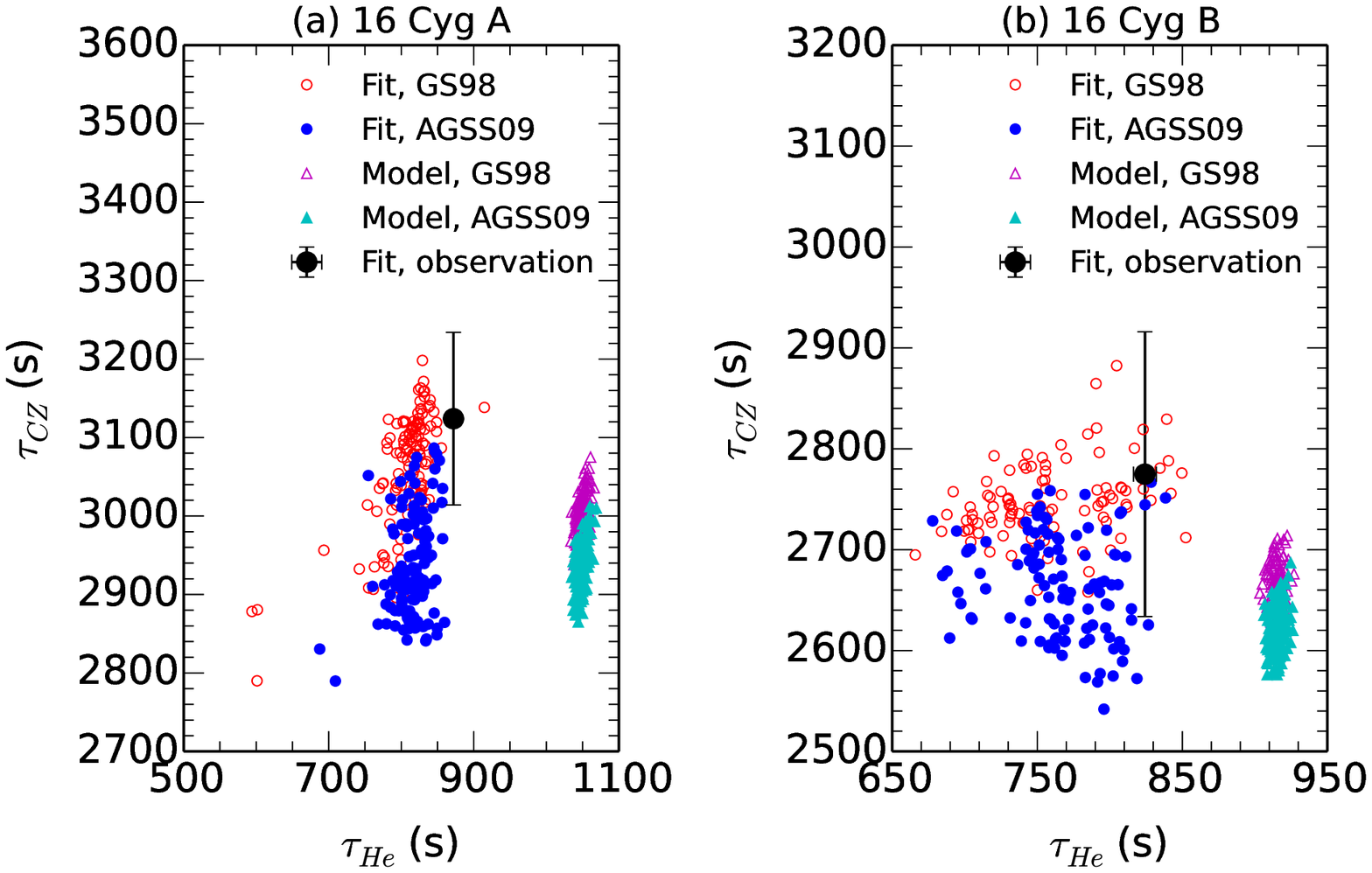}
\plotone{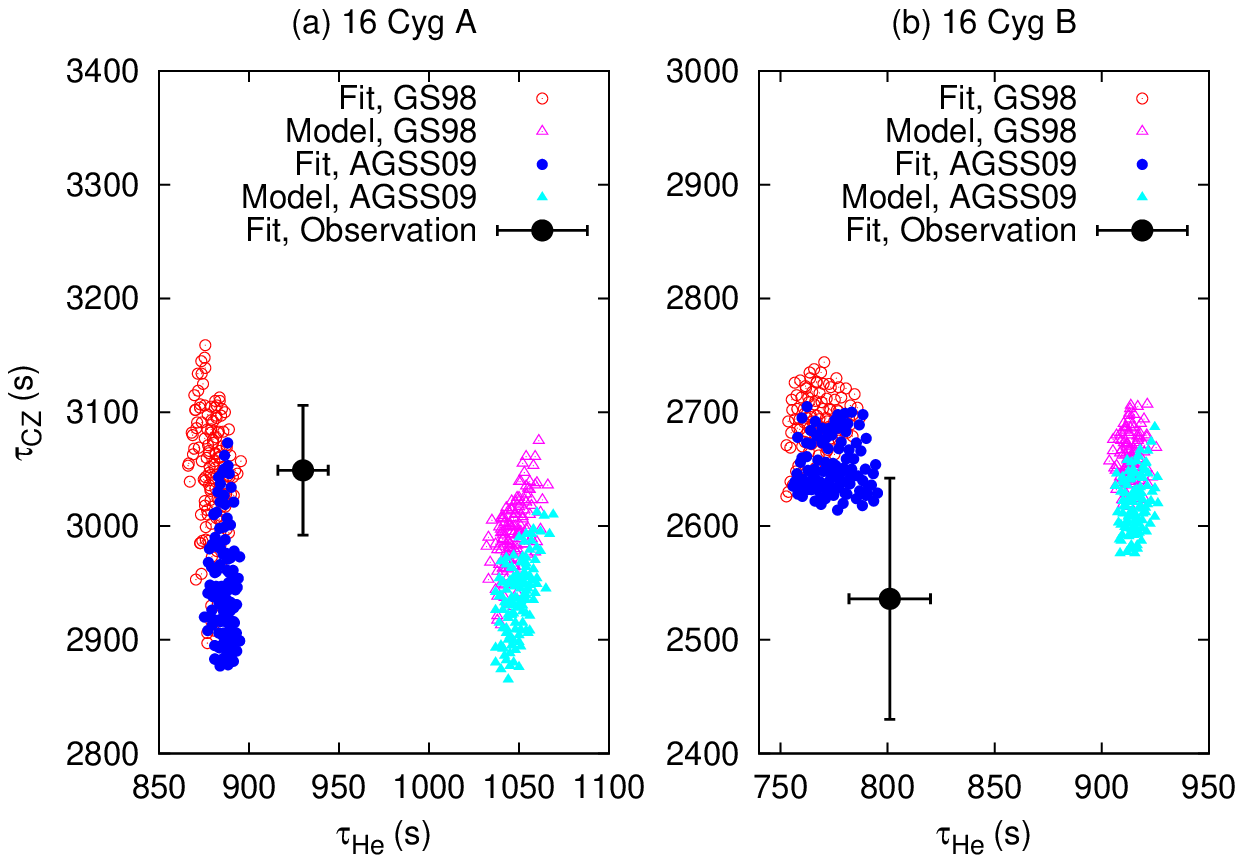}
\caption{Comparing $\tau_\mathrm{He}$ and $\tau_\mathrm{CZ}$ for (a) \SCA\ and (b) 
\SCB. The black point with error-bars represent the observations. 
Other points represent the $\tau$ values for the MESA models. 
The top panels show the results using Method A, the middle panels show those
using Method B and the bottom panels show the same using Method C.}
\label{fig:tau}
\end{figure}

The oscillatory signal in the observed frequencies contains information about the
location of the acoustic glitches. Fig.~\ref{fig:tau} shows the acoustic depths of 
the base of the convection zone and the HeII ionization zone obtained using all 3 methods.
The $\tau$ for the models were estimated in two different ways; 
one by calculating the acoustic depth from the known sound-speed profile of the 
model (Eq.~\ref{eq:tau}), and the other by fitting the second differences of the 
model frequencies (Eq.~\ref{eq:fit_eq_a}) or by fitting the model frequencies 
(Eq.~\ref{eq:fit_eq_b} or \ref{eq:fit_eq_c}).
The first estimate using Eq.~\ref{eq:tau}, is affected by the uncertainties
in the definition of stellar surface. This issue has been extensively discussed
by \citet{houd07} and will affect the acoustic depth of all layers by the
same amount. However, the uncertainties in $\tau$ do not affect the determination of $Y$ which depends
on the amplitude of the oscillatory signal.
As can be seen in the figure, for 
$\tau_\mathrm{CZ}$ the two estimates for a model match well, but for $\tau_\mathrm{He}$ 
the value obtained from Eq.~(\ref{eq:tau}) is systematically higher than that 
obtained by fitting the second differences or the frequencies. This systematic 
shift is present in the results of all the three fitting methods, and could be 
due to the fact that for $\tau_\mathrm{CZ}$ the location of the discontinuity 
in the sound-speed gradient is well-defined and there is little ambiguity in its 
position; for $\tau_\mathrm{He}$ on the other hand there is, strictly speaking, 
no real discontinuity at all, but merely a sharp variation in $\Gamma_1$ that 
gives rise to the oscillatory signal in the frequencies. While using Eq.~(\ref{eq:tau}) 
we used the minimum in $\Gamma_1$ in the HeII ionization zone as the location 
of the `discontinuous' layer, and it is possible that the effective location of 
the `discontinuity' is at a layer above this. The value of $\tau_\mathrm{CZ}$ 
obtained by fitting the observed frequencies is close to what we get for our 
models; $\tau_\mathrm{He}$ obtained from the observed frequencies lies within 
about $3\sigma$ of that obtained by fitting the frequencies of the models.  The 
amplitude of signal due to the base of the convection zone in the observed 
frequencies is consistent with those in the models' frequencies.

\begin{figure}
\epsscale{1.0}
\plotone{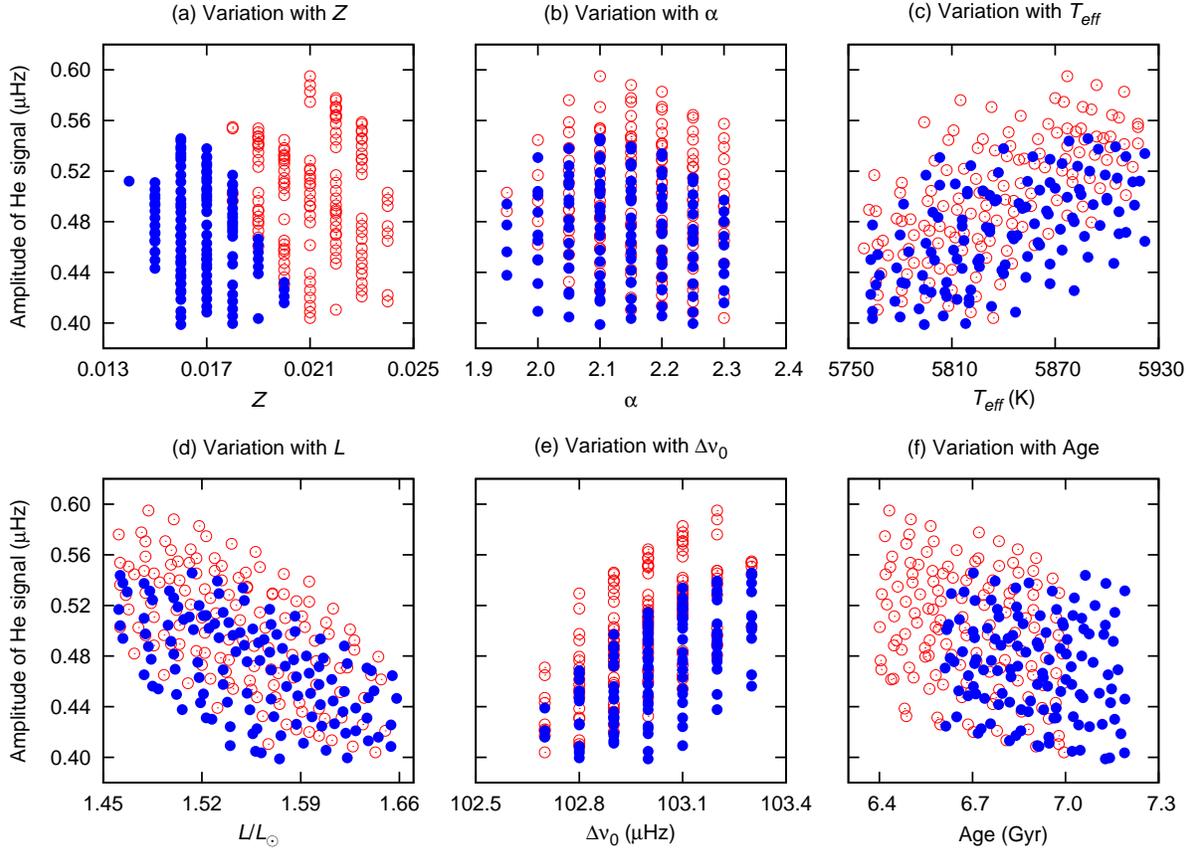}
\caption{The variation of the average amplitude of the helium signal as a function 
of different model parameters. Results are shown for \SCA\ MESA models and 
amplitudes are calculated using Method A. Models with both GS98 (red open circles) 
and AGSS09 (blue filled circles) heavy-element mixtures are shown.}
\label{fig:par}
\end{figure}

We examined the variation of the average amplitude of the HeII signal with the input 
parameters of the models and their seismic characteristics. Fig.~\ref{fig:par} shows 
the variation of the average amplitude calculated using the fitting Method A for \SCA. 
We can see that the average amplitude of the HeII signal is not particularly sensitive 
to the metallicity and mixing length parameter (see Fig.~\ref{fig:par}a, \ref{fig:par}b); 
however, it shows systematic variation as a function of effective temperature, luminosity, 
large frequency separation, and  age (see Fig.~\ref{fig:par}c-f). 
Much of the trend is essentially due to variation in $Y$ with these parameters rather 
than variation of the amplitude for a given $Y$. The vertical dispersion in each of 
these figures is due to the dispersion in mass, which also changes $Y$ at a given set 
of parameters. The variation of the amplitude was very similar when calculated using 
Method B and C.

\section{Conclusions}
\label{sec:con}

We have used the oscillatory signature in the oscillation frequencies of \SCAB\ 
caused by the depression of $\Gamma_1$ in the  HeII ionization zone to determine 
the current helium abundance and the depth of the HeII ionization zone. For this 
task we have used frequencies obtained with 2.5 years of {\it Kepler} observations. 
We have used three different techniques combined with three sets of stellar models 
for calibrating the signal. We have demonstrated that the helium abundance of 
these stars can be reliably determined using the observed frequencies. For \SCA\ 
the three methods give results that are consistent with the error bars, while 
for \SCB\ the results can differ, though the difference is only $2.5\sigma$ in 
the worst case. The helium abundance for \SCA\ is found to lie in the range 0.231 
to 0.251 and that of \SCB\ lies in the range 0.218 to 0.266. 

The error bars quoted in Table~\ref{tab:yresults} are the random errors arising from those
in the observed frequencies. In addition, there would be systematic uncertainties
due to approximate form of oscillatory term as well as uncertainties in the used stellar
models. The first contribution can be estimated from the differences in
values obtained using the three techniques and is already included in the values
quoted above. The stellar-model uncertainty should  be mainly from the
EOS which translates the helium abundance to $\Gamma_1$.
From extensive tests for the Sun, it is known that the OPAL EOS is close
to that of the Sun \citep[see e.g.][]{basu97} and hence we do 
not expect much uncertainty on that count in these stars where the uncertainties
in the frequencies are much larger than those for the Sun.

This technique for determining helium abundance is not particularly
sensitive to the presence of surface effect --- the surface corrections
are smooth function of $n$ and will mainly contribute to the smooth part
of the frequency and can be separated from the oscillatory part. This is particularly
true when the observed frequencies are available over a wide range of $n$ values
as is the case with both stars considered here. If the observed frequency
range is limited, there could be some difficulty in separating the helium
signal from smooth part since the oscillatory signal can also
be approximated with moderately high degree polynomial. This is the reason 
why we restrict the degree of polynomial in Method A to the minimum value 
that is statistically significant. The regularization used in Methods~B and C
allows us to use a more complex function to account for the smooth component.
\citet{basu04} have used similar technique to estimate the helium abundance
in the solar envelope using only low degree modes and find results which are consistent with the known
value obtained from more detailed seismic inversions.

The estimates of the helium mass fraction are those of the current abundance in 
the outer envelope of the stars. This value is, of course, lower than the initial 
helium abundance of both stars because of the gravitational settling of helium. 
We can determine the initial helium abundance from the current one by assuming 
that the models of the two stars correctly represent the amount of helium 
depleted due to settling. MESA models show a depletion of $0.048\pm0.004$ (for 
GS98 mixture) and $0.054\pm0.006$ (for AGSS09 mixture) for \SCA. Depletion of 
helium is lesser for \SCB\ models, $0.043\pm0.006$ for GS98 and $0.048\pm0.007$ 
for AGSS09 mixture. Since the two components of the binary system are believed 
to have formed from the same gas cloud, the initial composition should be the 
same for both stars. In that case the current helium abundance in the envelope 
of \SCB\ should be higher than that in \SCA\ by about 0.005. The values that we 
find are consistent with this difference. This difference arises because of 
small difference in masses of the two stars. Similarly, the difference between 
GS98 and AGSS09 mixtures arise due to the difference in the thickness of convection 
zones in the two cases. Since GS98 models have deeper convection zone, the 
depletion of helium is lower. Thus the initial helium abundance of the \SC\ system
is between 0.28 and 0.31. This 
is somewhat larger than the value found by \citet{metc12} and close to the values 
found by \citet{grub13}. The solar initial helium abundance was determined to 
be $0.278 \pm 0.006$ \citep{sere10}. Thus our estimates of $Y_0$ for the \SC\ 
system is consistent with solar values particularly considering the fact that 
these stars have higher metallicity.

The fitted value of $\tau_\mathrm{He}$ obtained from observed frequencies for 
\SCB\ matched that obtained from fitting the model frequencies, but for \SCA\ 
the observed value is larger than that in all selected models. This could be due 
to some systematic errors in modelling or in the other observed quantities that 
were used to constrain the models.

The two sets of MESA models constructed with different heavy-element mixtures 
were used to check if the observed frequencies could distinguish between the 
different mixtures. For both sets of models, the value of $Y$, as well as 
$\tau_\mathrm{He}$ and $\tau_\mathrm{CZ}$ are very similar, although, as expected, 
there is a difference in the depth of the convection zone between the two sets 
with  models constructed  with the AGSS09 having a shallower convection zone 
depth and hence lower $\tau_\mathrm{CZ}$ than models with the GS98 mixture. 
But there is considerable overlap in the range for each set of models. For \SCB\ 
the uncertainty in the value of $\tau_\mathrm{CZ}$ obtained from observed 
frequencies is too large to distinguish between the two sets of models. For \SCA\ 
the error in $\tau_\mathrm{CZ}$ from observed frequencies is reasonably small 
and if the systematic errors in modelling can be sorted out, it may be possible 
to use the models to distinguish between the two sets of models and get an 
independent handle on the controversy over the solar heavy-element abundance.

\acknowledgments
SB acknowledges support from NSF grant AST-1105930 and NASA grant NNX13AE70G.
AM acknowledges support from the NIUS programme of HBCSE (TIFR).
MJPFGM and JPF acknowledge support from the EC Project SPACEINN ({\small FP7-SPACE-2012-312844}) and
JPF was also supported by the European Research Council (EC-FP7) through a fellowship of the Starting 
Grant {\small ERC-2009-StG-239953}. WJC acknowledges support from the UK Science and Technology 
Facilities Council (STFC). Funding for the Stellar Astrophysics Centre (SAC) is provided by The
Danish National Research Foundation.
\ \ \

%

\begin{deluxetable}{cccc}
\tabletypesize{\small}
\tablecaption{Various observables used to constrain the models of 16~Cyg A and B.}
\tablewidth{0pt}
\tablehead{\colhead{Observable} & \colhead{\SCA} & \colhead{\SCB} & \colhead{Reference}}
\startdata
$T_\mathrm{eff}$ & 5839$\pm$42 K & 5809$\pm$39 K & 1 \\
$\log g$ & 4.29$\pm$0.02 dex & 4.36$\pm$0.02 dex & 1 \\
$L$ & 1.56$\pm$0.05 $L_\odot$ & 1.27$\pm$0.04 $L_\odot$ & 2 \\
\febyh & 0.096$\pm$0.040 dex & 0.096$\pm$0.040 dex & 3 \\
$\Delta_0$ & 102.9$\pm$0.2 $\mu$Hz & 116.5$\pm$0.2 $\mu$Hz & \nodata \\
$\langle d_{02}\rangle$ & 5.82$\pm$0.03 $\mu$Hz & 6.70$\pm$0.03 $\mu$Hz & \nodata 
\enddata
\tablecomments{Surface \febyh\ is assumed to be same for both components with increased 
uncertainty. Average large and small separations were calculated using observed 
frequencies.}
\tablerefs{(1) \citet{whit13}; (2) \citet{metc12}; (3) \citet{rami09}}
\label{tab:obs}
\end{deluxetable}

\begin{deluxetable}{cccccccccc}
\tabletypesize{\scriptsize}
\rotate
\tablecaption{Observed oscillation frequencies for 16~Cyg A and B.}
\tablewidth{0pt}
\tablehead{
& \multicolumn{4}{c}{16~Cyg A} & &\multicolumn{4}{c}{16~Cyg B} \\
\cline{2-5} \cline{7-10} \\
\colhead{n} &\colhead{$l = 0$} &\colhead{$l = 1$} &
\colhead{$l = 2$} &\colhead{$l = 3$} & &\colhead{$l = 0$} &
\colhead{$l = 1$} &\colhead{$l = 2$} &\colhead{$l = 3$} \\
& \colhead{($\mu$Hz)} &\colhead{($\mu$Hz)} &\colhead{($\mu$Hz)} &
\colhead{($\mu$Hz)} & &\colhead{($\mu$Hz)} &\colhead{($\mu$Hz)} &
\colhead{($\mu$Hz)} &\colhead{($\mu$Hz)} 
}
\startdata
11 &        \nodata     & 1334.392$\pm$0.034 & 1384.368$\pm$0.040 &        \nodata     &
&           \nodata     &        \nodata     &        \nodata     &        \nodata     \\
12 & 1391.639$\pm$0.037 & 1437.520$\pm$0.043 & 1488.353$\pm$0.100 &        \nodata     &
&           \nodata     &        \nodata     & 1686.327$\pm$0.057 &        \nodata     \\
13 & 1494.996$\pm$0.058 & 1541.914$\pm$0.050 & 1591.242$\pm$0.115 &        \nodata     &
&    1695.066$\pm$0.056 & 1749.184$\pm$0.050 & 1804.197$\pm$0.235 &        \nodata     \\
14 & 1598.695$\pm$0.066 & 1645.058$\pm$0.095 & 1694.193$\pm$0.175 &        \nodata     &
&    1812.420$\pm$0.094 & 1866.525$\pm$0.102 & 1921.176$\pm$0.168 &        \nodata     \\
15 & 1700.909$\pm$0.083 & 1747.154$\pm$0.082 & 1795.757$\pm$0.105 & 1838.283$\pm$0.355 &
&    1928.900$\pm$0.068 & 1982.587$\pm$0.073 & 2036.677$\pm$0.123 & 2085.517$\pm$0.260 \\ 
16 & 1802.316$\pm$0.068 & 1848.980$\pm$0.054 & 1898.284$\pm$0.100 & 1940.756$\pm$0.413 &
&    2044.274$\pm$0.058 & 2098.076$\pm$0.056 & 2152.405$\pm$0.100 &        \nodata     \\
17 & 1904.611$\pm$0.054 & 1951.997$\pm$0.050 & 2001.647$\pm$0.070 & 2045.922$\pm$0.229 &
&    2159.580$\pm$0.060 & 2214.159$\pm$0.058 & 2268.966$\pm$0.082 & 2319.208$\pm$0.225 \\
18 & 2007.572$\pm$0.045 & 2055.527$\pm$0.046 & 2105.314$\pm$0.051 & 2149.914$\pm$0.143 &
&    2275.946$\pm$0.045 & 2331.138$\pm$0.044 & 2386.267$\pm$0.060 & 2436.792$\pm$0.176 \\
19 & 2110.915$\pm$0.040 & 2159.153$\pm$0.045 & 2208.907$\pm$0.067 & 2253.539$\pm$0.172 &
&    2392.718$\pm$0.040 & 2448.250$\pm$0.042 & 2503.575$\pm$0.061 & 2554.149$\pm$0.150 \\
20 & 2214.226$\pm$0.049 & 2262.536$\pm$0.047 & 2312.545$\pm$0.091 & 2357.368$\pm$0.181 &
&    2509.668$\pm$0.038 & 2565.398$\pm$0.042 & 2620.566$\pm$0.064 & 2671.754$\pm$0.155 \\
21 & 2317.326$\pm$0.056 & 2366.260$\pm$0.059 & 2416.301$\pm$0.137 & 2462.077$\pm$0.373 &
&    2626.399$\pm$0.042 & 2682.409$\pm$0.048 & 2737.741$\pm$0.081 & 2789.187$\pm$0.270 \\
22 & 2420.931$\pm$0.089 & 2470.322$\pm$0.081 & 2520.695$\pm$0.266 &        \nodata     &
&    2743.327$\pm$0.061 & 2799.735$\pm$0.062 & 2855.682$\pm$0.137 & 2906.915$\pm$0.371 \\
23 & 2525.141$\pm$0.197 & 2574.639$\pm$0.138 & 2624.147$\pm$0.366 & 2669.376$\pm$0.869 &
&    2860.773$\pm$0.110 & 2917.799$\pm$0.100 & 2973.470$\pm$0.248 & 3024.940$\pm$0.941 \\
24 & 2629.212$\pm$0.212 & 2679.951$\pm$0.188 & 2727.410$\pm$1.643 &        \nodata     &
&    2978.481$\pm$0.162 & 3036.084$\pm$0.156 & 3093.204$\pm$0.517 & 3144.365$\pm$0.679 \\
25 & 2733.199$\pm$0.470 & 2784.178$\pm$0.369 &        \nodata     &        \nodata     &
&    3097.575$\pm$0.492 & 3154.418$\pm$0.271 &        \nodata     &        \nodata     \\
26 & 2837.442$\pm$0.682 & 2891.303$\pm$0.841 &        \nodata     &        \nodata     &
&    3214.348$\pm$0.562 & 3272.941$\pm$0.741 &        \nodata     &        \nodata     \\
27 & 2944.587$\pm$1.226 & 2996.533$\pm$1.022 &        \nodata     &        \nodata     &
&           \nodata     &        \nodata     &        \nodata     &        \nodata   
\enddata
\label{tab:freq}
\end{deluxetable}

\begin{deluxetable}{ccccl}
\tabletypesize{\small}
\tablewidth{0pt}
\tablecaption{Physical parameters obtained by fitting the observed frequencies 
of 16~Cyg A and B.}
\tablehead{
\colhead{Method}&\colhead{Amplitude of CZ}&\colhead{$\tau_\mathrm{CZ}$}
&\colhead{Amplitude of He}&\colhead{$\tau_\mathrm{He}$}\\
\colhead{}&\colhead{Signal ($\mu$Hz)}&\colhead{(s)}&\colhead{Signal ($\mu$Hz)}&\colhead{(s)}
}
\startdata
\multicolumn{5}{c}{16~Cyg A}\\
A& $0.072\pm0.011$& $3079\pm54\ph$& $0.492\pm0.013$& $919\pm9\ph$\\
B& $0.030\pm0.012$& $3124\pm110$& $0.435\pm0.036$& $872\pm4$\\
C& $0.055\pm0.012$& $3049\pm57\ph$& $0.508\pm0.017$& $930\pm14$\\[5pt]
\multicolumn{5}{c}{16~Cyg B}\\
A& $0.043\pm0.010$& $2552\pm147$& $0.421\pm0.017$& $820\pm17$\\
B& $0.048\pm0.018$& $2775\pm141$& $0.900\pm0.031$& $824\pm8$\\
C& $0.044\pm0.012$& $2536\pm106$& $0.449\pm0.018$& $801\pm19$
\enddata
\label{tab:fitpar}
\end{deluxetable}

\begin{deluxetable}{cccc}
\tabletypesize{\small}
\tablewidth{0pt}
\tablecaption{Helium abundances of 16~Cyg A and B.}
\tablehead{
\colhead{Method} &\multicolumn{2}{c}{MESA} &\colhead{YREC}\\
\colhead{} &\colhead{GS98} &\colhead{AGSS09} &\colhead{GS98}
}
\startdata
\multicolumn{4}{c}{16~Cyg A}\\
A& $0.238\pm0.009$& $0.243\pm0.009$& $0.231\pm0.009$\\
B& $0.239\pm0.021$& $0.242\pm0.023$& $0.236\pm0.016$\\
C& $0.250\pm0.009$& $0.251\pm0.009$& $0.249\pm0.009$\\[5pt]
\multicolumn{4}{c}{16~Cyg B}\\
A& $0.263\pm0.012$& $0.266\pm0.012$& $0.257\pm0.009$\\
B& $0.218\pm0.013$& $0.228\pm0.011$& $0.219\pm0.009$\\
C& $0.251\pm0.010$& $0.254\pm0.010$& $0.255\pm0.009$
\enddata
\label{tab:yresults}
\end{deluxetable}

\end{document}